\documentclass[12pt]{elsarticle}
\makeatletter
\def\ps@pprintTitle{%
 \let\@oddhead\@empty
 \let\@evenhead\@empty
 \def\@oddfoot{}%
 \let\@evenfoot\@oddfoot}
\makeatother
\usepackage{graphicx}
\usepackage{tabularx}
\usepackage{amsmath}
\graphicspath{{images/}}

\begin{document}

\begin{frontmatter}

\title{A Network Based Approach to Characterize  Twenty-First-Century Populism in Colombia}

\author[1]{Juan D. Garcia-Arteaga}
\ead{judgarciaar@unal.edu.co}
\author[2]{Valentina Pellegrino}

\address[1]{School of Medicine,\\ Universidad Nacional de Colombia, \\Bogotá, Colombia}

\address[2]{LASC: Laboratory for the Anthroplogy of the State in Colombia, \\Bogotá, Colombia}

\begin{abstract}
{Populism is a political phenomenon of democratic illiberalism centered on the figure of a strong leader. By modeling  person/node connections of prominent figures of the recent Colombian political landscape  we map, quantify, and analyze the position and influence of Alvaro Uribe as a populist leader. We found that Uribe is a central hub in the political alliances networks, cutting through traditional party alliances,  
but is not the most central figure in the state machinery.
The article first presents the framing of the problem, followed by the historical context of the case in study, the methodology employed and data collection, analysis, conclusions and further research paths. This study has implications for offering a new way of applying quantitative methods to the studies of populist regimes.}


\end{abstract}
\begin{keyword}
 Populism \sep Political \sep Networks \sep Colombia
\end{keyword}

\end{frontmatter}

\section{Introduction}
Populism is a phenomenon broadly defined as the pursuit of politicial power by use of a confontational narrative founded on the concept of ``us, the people'' against ``them, the elite''. Once considered a result of politcally immature systems, the rise of populist figures in well established liberal democracies in Europe and North America\cite{pappas2016modern}  has renewed interest in the subject  and in understanding how and why these regimes come to power  \cite{de2018democraduras}\cite{kaltwasser2017oxford}.

Although populist focused studies date back to the 1960's,  there is still not a consensus on whether populism should be defined as a communication style, an ideology  or a political strategy.  The research question in contemporary studies has then shifted from what populism is to how may populism  be quantified. Recently the field has added studies about the degree of populism in specific contexts, recurring to textual and content analysis of discourses, speeches or manifestos \cite{hawkins2019measuring}\cite{jagers2007populism}\cite{pauwels2011measuring}\cite{popping2018measuring}. Additionally, there has been an increase in the studies of the attitudes in citizens that can be supportive of populist regimes \cite{cisneros2018voto} \cite{akkerman2014populist} \cite{rooduijn2018unites} \cite{hawkins2012measuring}. 

These studies are valuable as there is much gain in understanding the discourse patterns of populist leaders and how masses are incorporated into them.  However, these approaches tend to ignore both the impact of populism in the party system \cite{hawkins2017ideational} and, more generally, how populists leaders fit in in the landscape of political power relationships.  
 It is possible that, by centering in the figure of the leader and his or her discourse, we are falling into the trap of methodological individualism denounced by \cite{victor2017introductionPoliticalNetworks}: a tendency to focus on individuals instead of relationships, which entails unrealistic assumptions of the independence of these individuals.

  In this article we present an alternative to a widespread tendency to study populism and the networks that support it separately \cite{roberts2017populism}. Specifically, we will analyze the network of political alliances and work relationships surrounding a populist leader: Colombia's former president Alvaro Uribe whose ongoing influence and politics have been used as an example of  neo-populism by various authors \cite{nasi2007derechizacion,fierro2014alvaro}.

   We offer a novel analysis based on  the Colombian political networks to show how Uribe has a central role in them, contesting the commonly held belief of the source of populist power being the direct contact of the leader with the masses without the mediation of traditional political machinery.  Our research is based on  publicly available profiles documenting different relationships types such as political alliances, work relationships, rivalries and family ties of the main actors of Colombian  politics.

   The article is organized as follows. In the rest of the Introduction we  provide a general context of Colombian politics in the last two decades and  briefly review the use of networks in the analysis of political problems. We then describe the dataset used to capture the relationships between different actor of the Colombian political scene. The data collected is then analyzed using well known concepts of network analysis such as centrality and community detection. We show how these concepts relate to political strategies  put in place to consolidate power. 
   
\subsection{Alvaro Uribe and the Colombian Political Landscape}

A former congressman, Governor and Mayor of Medellín, Alvaro Uribe distanced himself from the Liberal Party because of his hard-line security politics. Running as an independent in the 2002 presidential elections and winning by a landslide, Uribe became the first president in Colombian history outside of the traditional two-party system.  Once in office, he used his immense popularity and political influence  to change the constitution and remove the ban on presidential reelection, thus allowing himself to be reelected  for a second term (2006-2010). 

Unable to run for a third term, Uribe supported the candidacy of Juan Manuel Santos, his former minister of Defense and assumed ally. Santos was elected for the 2010-2014 term by cashing in Uribe's electoral force and was re-elected for the 2014-2018 term on a completely different platform  based on signing a peace treaty with FARC guerrilas.  Uribe then became the most vocal and visible opponent to the peace project and to the  Santos administration, creating a right wing opposition party, \textit{Centro Democrático} (Democratic Center, hereafter CD), and being elected as a Senator in the Colombian Congress twice (2014-2018 and 2018-2022).

The CD now constitutes the majority of the Congress. Colombia's current president (Iván Duque, 2018-2022) belongs to the CD  and was elected under the strong support and campaigning of Uribe.   

Uribe  is  one of the most interesting examples of neo-populism as, on the one hand,  he and his communcation style fulfill the commonly accepted  characteristics of a populist leader:  charismatic,   mobilizes masses appealing to a direct communication style,  presents himself as an outsider of the traditional politics,  promotes an ``us against them'' adversarial politics that minimizes the space for democratic deliberation, and  promotes the idea of his manifest destiny as the only possible ``saviour'' of the country from its perceived enemies. On the other hand, his status as populist has been challenged mainly because his origins are strongly rooted in the traditional Colombian economic and political elite, undermining the ``people vs the elites'' opposition that characterizes many populist regimes.

\subsection{Individual Power Networks}

There is a widespread structuralist approach in the social sciences which defines power as relational rather than inherent to the agent \cite{knoke1994political}, \textit{i.e.} actor $A$'s power is defined in relation to his influence  over actor $B$.  Political science  may then be seen as the study of the relationships that create, access, use, accumulate and/or preserve power within a society and politics themselves may be understood fundamentally as a network phenomenon\cite{mcclurg2014political}.  This relational approach is  slowly gaining foothold in a discipline mostly grounded on theories that  assume independence of actors.

The use of networks as an adaptable tool to describe various relationships between individuals, as opposed to other collective actors such as states or political parties, has been adopted by various researchers to describe political data\cite{victor2016oxford}\cite{knoke1994political}. An important part of this research is based on the analysis of digital data from social networks and  focuses on individuals and potential voters. The relatively immediate availability of this type of data has promoted a large corpus of research describing the formation of communities with similar political points of view,  the relationship between political leader and their followers  and predicting the political alignment  of potential voters \cite{conover2011predicting,colleoni2014echo}.

The research of networks  the inner structures of power wielding political instituions are rarer, possibly due to the difficulty of accessing descriptive data \cite{siegel2011social}. In some cases, relational data may be manually collected from public information \cite{moniz2016empirical, gil1996political} or from direct surveys \cite{asif2016connections,huhe2018evolution}, but the process is work intensive, slow and difficult to keep up to date \cite{szwarcberg2012revisiting}.  

Affiliation networks \cite{breiger1974duality, lattanzi2009affiliation}, bipartite graphs in which one set corresponds to individuals and the other to groups to which the individuals belong, have been used as an alternative to direct relationships. When available, it is possible to use groups  corresponding to formal and well defined affiliations, \textit{e.g.} families \cite{cruz2017politician} or political parties \cite{faustino2019data}. In other cases, especially in political systems without a strong culture of information openness, authors have to use  more flexible relation definitions to build the networks, \textit{e.g.} alumni-connections of US congressmen \cite{battaglini2019effectiveness} or intelligence reports of the simultaneous presence in public events of Soviet Politicians during     the Breshnev era \cite{faust2002scaling}.

Relationships between members of deliberative assemblies, such as parliaments and legislative bodies, may be uncovered  by analyzing the coincidences of their voting pattern in support of bills \cite{Signorelli_2017,parigi2014political,fischer2019mps}.

In the following Section we will describe a semi-automatic method to collect direct relational data (``works with'', ``is allied with'') of prominent figures of recent Colombian history.
\section{Data}

Data was collected from the \textit{``Quien es quien''} (``Who is Who'') section of  \textit{``La Silla Vacia''} (``The Empty Chair''), an independent  Colombian news and political journalism website \footnote{\texttt{http://lasillavacia.com/quienesquien}} using web scrapping tools implemented in the Python 3 programming language \cite{10.5555/1593511}. The raw data consisted on 344 individual profiles saved as plain html text webpages.

\begin{figure}
  \centering
   \includegraphics[height=0.45\textwidth]{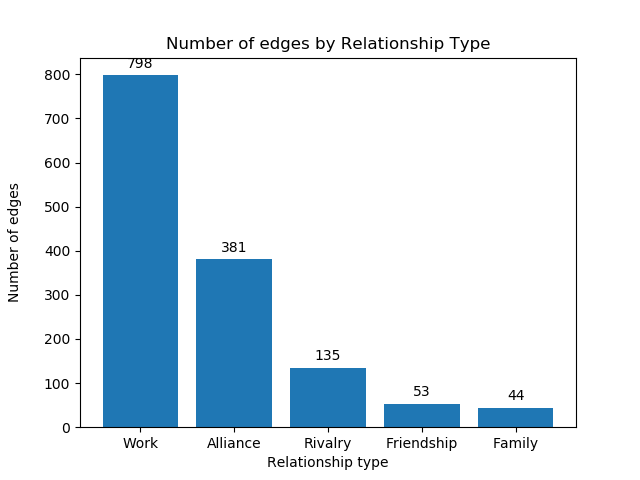}
   \caption{ \label{FIG:TipoDeEnlace}   Number of edges per relationship type.
}
  \end{figure}

  The webpage of each person in the network includes a brief profile and hyperlinks to related profiles. There are five types of relationships between profiles: Work, alliance, friendship, family and rivalry.   These relationships are, in general, non-exclusive, meaning two nodes may be related by more than one type of edge. The total number of edges by relationship type is presented in Figure \ref{FIG:TipoDeEnlace}. 

  It is clear from the figure that the most common type of relationship is ``Work'', with more edges (783) than all the other edge types combined (613), followed by ``Alliance'' (381).
 Although all edge information  could be represented either by a multiplex network, \textit{i.e.} one in which  each individual/node is connected by different edge/relationships \cite{breiger1986cumulated},  the difference in size would bias results towards ``Work'', the most populated edge type. 
 
\begin{figure}
  \centering
   \includegraphics[height=0.45\textwidth]{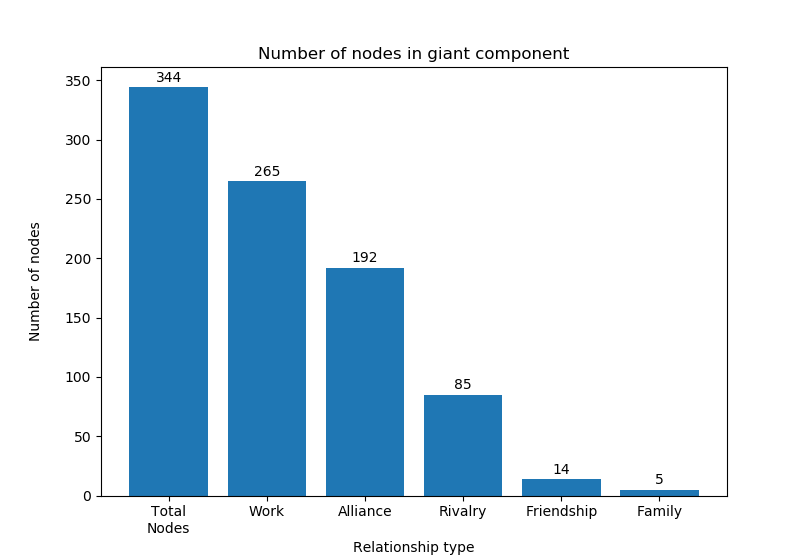}
   \caption{ \label{FIG:GC} Number of nodes in the largest connected component per relationship type graph.}
  \end{figure}

 We have opted  instead to analyze separately only the ``Work'' network  and ``Alliance'' networks (hereafter WN and AN, respectively) as they are the only edge types with a significant number of edges with respect to the number of nodes. Additionally, in both of the chosen networks there is a giant connected component (one in which a path exists between any pair of nodes) connecting more than half of the networks elements (Figure \ref{FIG:GC}). 
 
\section{Analysis}
\subsection{Degree }
\begin{figure}
  \centering
\begin{tabular}{cc}
   \includegraphics[width=0.495\textwidth]{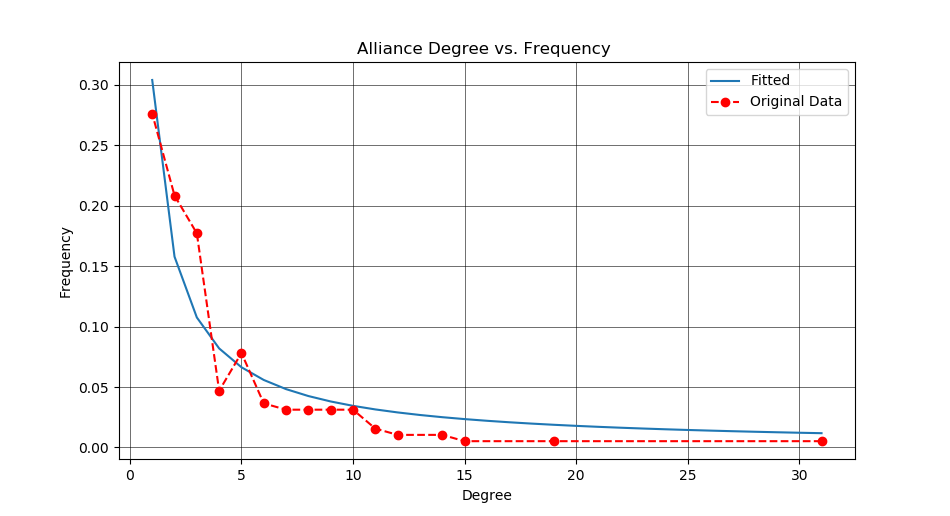}     &     \includegraphics[width=0.495\textwidth]{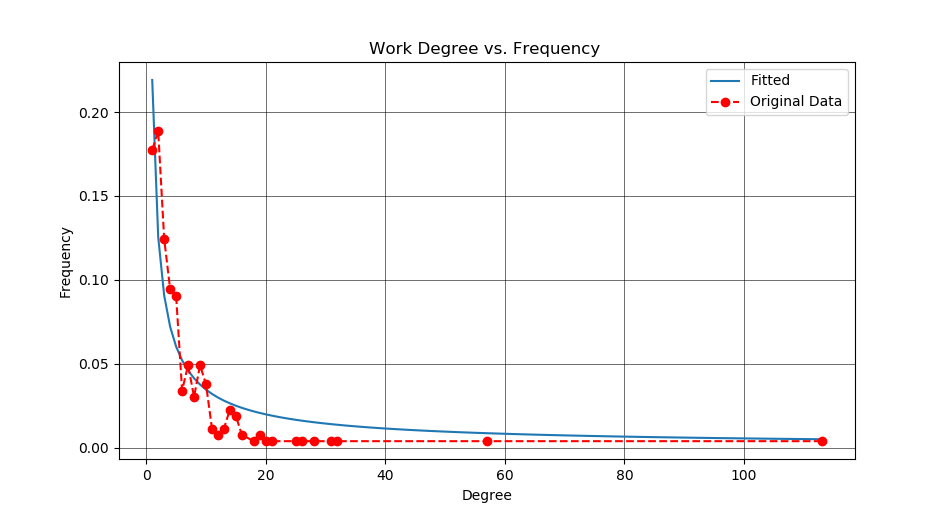}
\\
   \includegraphics[width=0.495\textwidth]{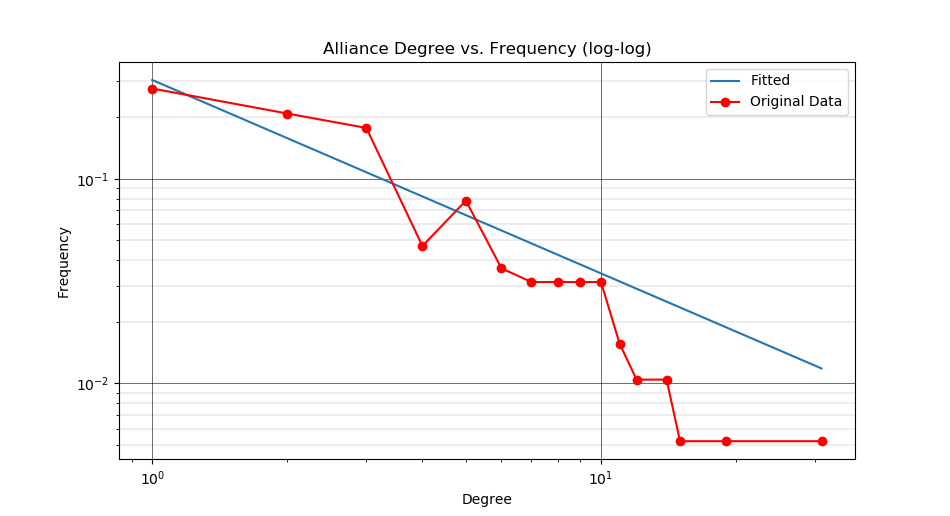}
     & 
        \includegraphics[width=0.495\textwidth]{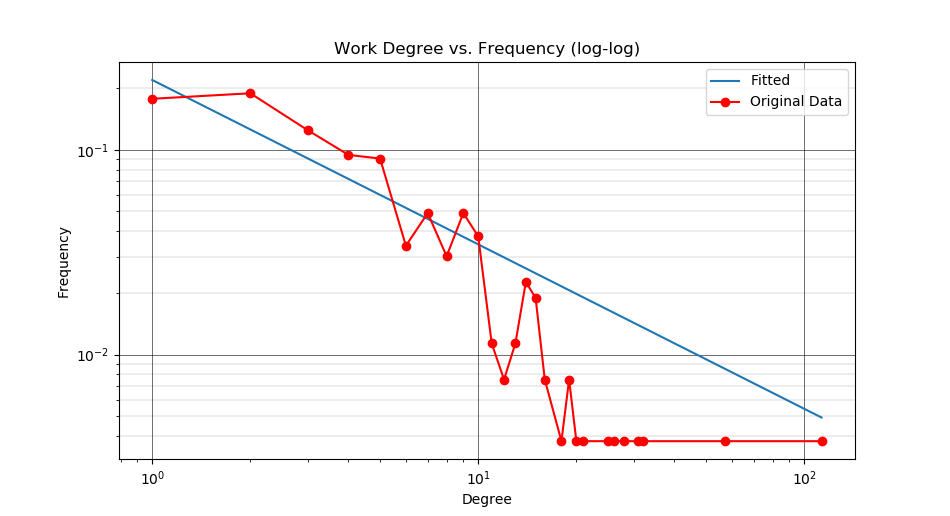}

\end{tabular}

   \caption{ \label{FIG:PowerLaw} Frequecies of node degrees for  AN (left) and WN (right). The fitted power law curves using  $\gamma$ values of $0.95$ for AN and $0.85$ for WN are overlain. Bottom row show the graphics in a logarithmic scale.}
  \end{figure}

Figure \ref{FIG:PowerLaw} shows the normalized frequencies of node degree (number of edges connected to each vertex) of WN and AN. Both networks show a  very long-tail and  nodes with degrees an order of magnitude above the average,  characteristics of power law distributions \cite{clauset2009power,haruna2019ordinal}. 

Formally, a distribution is said to follow a power law  if the fraction of nodes of degree $k$  is proportional to an exponential function:
\begin{equation}
P(k)\sim k^{-\gamma}.
 \end{equation}
  
Instances of natural, social and man-made networks spontaneously showing this behaviour have been extensively reported in the literature\cite{clauset2009power}.

Although there is still no consensus over the exact mechanisms  which result in this type of distribution, preferential attachment, also referred to as a ``Yule Process''\cite{yule1925ii} or ``the rich get richer effect'', has been proposed as a possible explanation. In this type of processes,  new edges  added to the network by a random or partly random process will attach with a higher probability to nodes with higher degrees in a self reinforcing loop\cite{albert2002statistical}.

\subsection{Centrality Measures}

A widely accepted, although loose, definition of a node's centrality coming from the study of social networks and organizations states that centrality is related to the ``importance'' of the node within the graph \cite{freeman1978centrality}. The most central nodes of AN and WN networks were calculated using four common  measures: Degree, Betweenness, Closeness and Eigenvector Centrality. By all four metrics Alvaro Uribe is the most central node of the AN and former President Juan Manuel Santos the most central node of the WN network. Results are shown in Table \ref{table:centAlianza}.

\begin{table}[t]
    \centering
\begin{tabular}{|cc||cc||cc||cc|}
\multicolumn{8}{c}{Alliance Network}\\\hline

\multicolumn{2}{|c||}{Degree}&\multicolumn{2}{c||}{Betweenness}&\multicolumn{2}{c||}{Closeness}&\multicolumn{2}{c|}{Eigencentrality}\\\hline
AU &  31& AU &  0.387& AU & 0.378& AU &  0.288\\
GV & 19 & EP & 0.187&  EP & 0.376 &EP & 0.276\\
\hline
\multicolumn{8}{c}{}\\

\multicolumn{8}{c}{Work Network}\\\hline
\multicolumn{2}{|c||}{Degree}&\multicolumn{2}{c||}{Betweenness}&\multicolumn{2}{c||}{Closeness}&\multicolumn{2}{c|}{Eigencentrality}\\\hline
JMS &  113 & JMS &  0.530 & JMS & 0.571 & JMS &  0.489\\
AU & 57 & AU & 0.203& AU & 0.466 & AU & 0.196\\
\hline
\end{tabular}
       \caption{\label{table:centAlianza}Top  centrality nodes of  AN and WN according to four metrics. In all measures Alvaro Uribe (AU) is the most central node of AN and the second most important of the WN. Former President Juan Manuel Santos (JMS) is the most important node by all measures of the WN. Other nodes are Germán Vargas Lleras (GV,  Vice President under Juan Manuel Santos)) and  Enrique Peñalosa (EP, former Mayor of Bogota).}
    \label{tab:my_label}

\end{table}

\subsubsection{AN Centrality}
It is interesting to note that, on the one hand, there is a large proportional gap between Alvaro Uribe's score in Degree and Betweenness measurements and the second most central  nodes.

Degree centrality (DC) is a simple yet effective way of measuring the importance of a node and corresponds to the number of edges connected to it.

Betweeness centrality (BC) assumes that information travels from node to node following the shortest path\cite{estrada2015first}. The BC of a node is  defined as:
\begin{equation}
BC(i)=\sum_i\sum_j \frac{\rho(i,j,k)}{\rho(j,k)}, i\neq j \neq k 
\end{equation}
where $\rho(j,k)$ is the number of shortest paths connecting nodes $i$ and $k$ and $\rho(j,i,k)$ is the number of shortest paths connecting $j$ and $k$
 passing through $i$.

\begin{figure}
  \centering
\begin{tabular}{c}
        \includegraphics[width=0.95\textwidth]{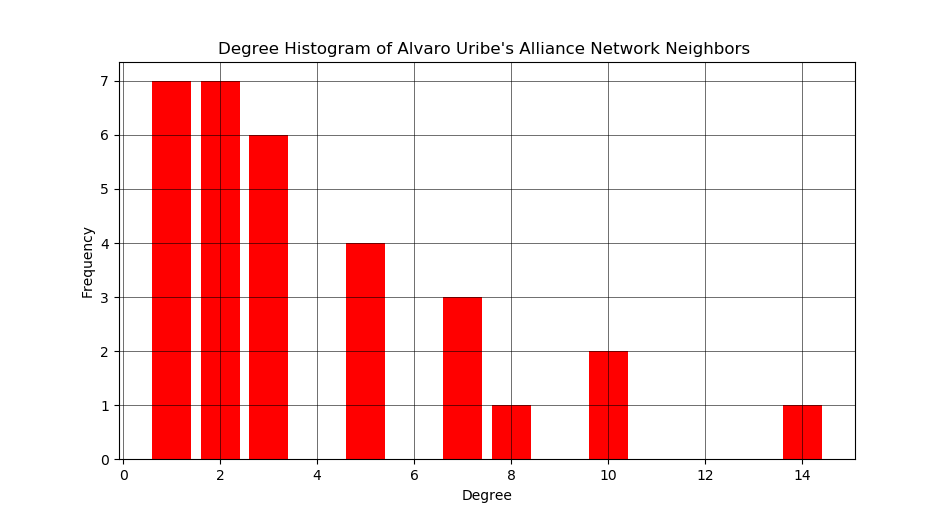}

\end{tabular}
   \caption{ \label{FIG:HistoUribe} DC histogram for the immediate neighbors of the Alvaro Uribe node in the AN. Most of the nodes connected to Alvaro Uribe have a very low degree.  }
  \end{figure}

A rapid inspection of the network shows that  Alvaro Uribe has many alliances with low-degree nodes, as showin in Figure \ref{FIG:HistoUribe}. Since DC measures the number of  edges connected to a node  (independently of the importance of the nodes it attaches with) and BC measures the number of shortest paths passing through a given node (again, independently of the importance of the paths), it is logical that having many connections, even to unimportant nodes, will boost these scores. 

On the other hand, Alvaro Uribe's Eigencentrality score (EC) is only marginally larger than other nodes. This measure is based on the concept that being connected to high-scoring nodes will contribute more to the centrality than being connected to low-scoring nodes, resulting in a more distributed centrality. The EC of a vertex $v_i$  in a graph $G$ is defined as:

\begin{equation}\label{eq:EC}
 EC(v_i)=
 {\frac {1}{\lambda }}\sum _{j\in G}a_{i,j}EC(v_j) 
\end{equation}
where $a_{i,j}$ is the value in row $i$ and column $j$ of $A$, the adjacency matrix of $G$. One may rewrite Equation \ref{eq:EC} as the eigenvector definition:
\begin{equation}\label{eq:Eigen}
A\mathbf{x}=\lambda \mathbf{x}.
\end{equation}

Variations of EC in which a node's centrality is determined by the centrality of its neighbors and  being connected to high-ranking nodes increases the influence in the network, are used to determine the ranking of webpages, \textit{e.g.} Google's  Pagerank algorithm \cite{page1999pagerank} or the citation impact of academic articles \cite{yan2009applying,bollen2006journal}. In the political context, it provides a strong incentive for actors with a limited political power to establish alliances with central nodes such as Uribe, thus increasing the latter's influence in a self perpetuating process consistent with the preferential attachment mechanism.

  \subsubsection{WN Centrality}

The WN is dominated by the figure of Juan Manuel Santos, who, as we previously explained, rose to the Presidency with the support of his former ally Uribe. Alvaro Uribe criticized Juan Manuel Santos's peace talk efforts, announced shortly after being sworn in, and has become one of his fiercest critics.

Besides Juan Manuel Santos and Alvaro Uribe, the top positions of the Work network are dominated either by former Presidents or by high-ranking ministers who have served under different Presidents. This, together with the difference in size of the AN and WN network (Figure \ref{FIG:GC}), tends to confirm the diagnostics of how, despite the changes stated in the 1991 constitution, Colombia remains a highly centralized nation with a large state-bureaucracy orbiting, mainly, around the executive branch of the government.
   \begin{figure}
  \centering
\begin{tabular}{c}
        \includegraphics[width=0.95\textwidth]{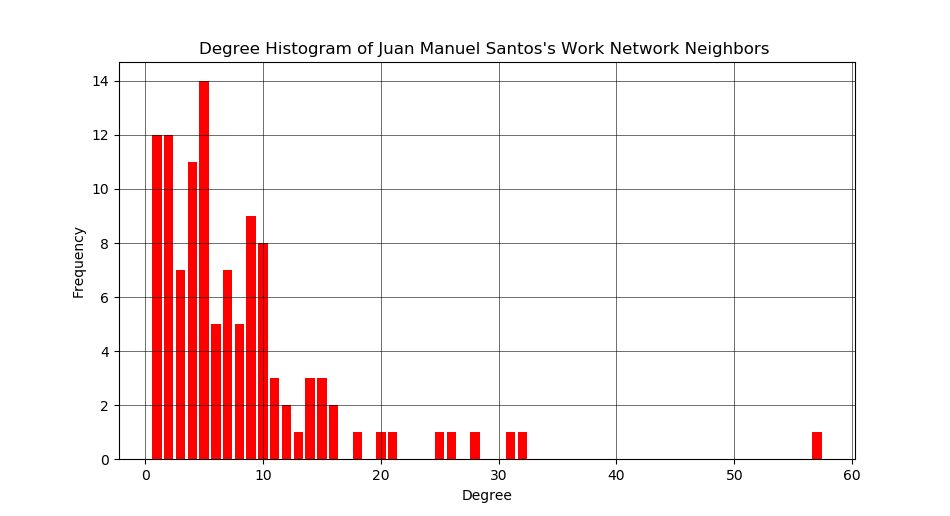}

\end{tabular}

   \caption{ \label{FIG:HistoSantos} DC histogram for the immediate neighbors of the Juan Manuel Santos node in the WN.  The highest connected node corresponds to Alvaro Uribe under whom Juan Manuel Santos served as a Minister of Defense. }
  \end{figure}

The higher centrality of Juan Manuel Santos over Alvaro Uribe in the WN is also consistent with their public discourses. Whereas Juan Manuel Santos insisted on the need to work towards peace construction and, consequently, created bureaucratic positions to negotiate and implement peace accords, Alvaro Uribe's discourse and public figure are centered on efficiency, action over reflection and the reduction of the size of the government, all points which align with the description of neo-populism given by Fierro \cite{fierro2014alvaro}.

It is interesting to note that one of the main criticisms of Alvaro Uribe and his party allies against Juan Manuel Santos is the alleged wide-spread use of pork barrel politics, \textit{i.e.} the appropriation of government spending for political gain. The use of key word repetition tactics by the CD on this point has made the Spanish word for ``sweet jam''    (\textit{``mermelada''}) a functional synonym of pork barrel in Colombia and popularized the  use of the expression ``spread the sweet jam''  (\textit{``repartir mermelada''}) to describe this practice \cite{angulo2019political}.

\subsection{Communities and Network Topology}

\begin{figure}
    \centering
\includegraphics[width=0.65\textwidth]{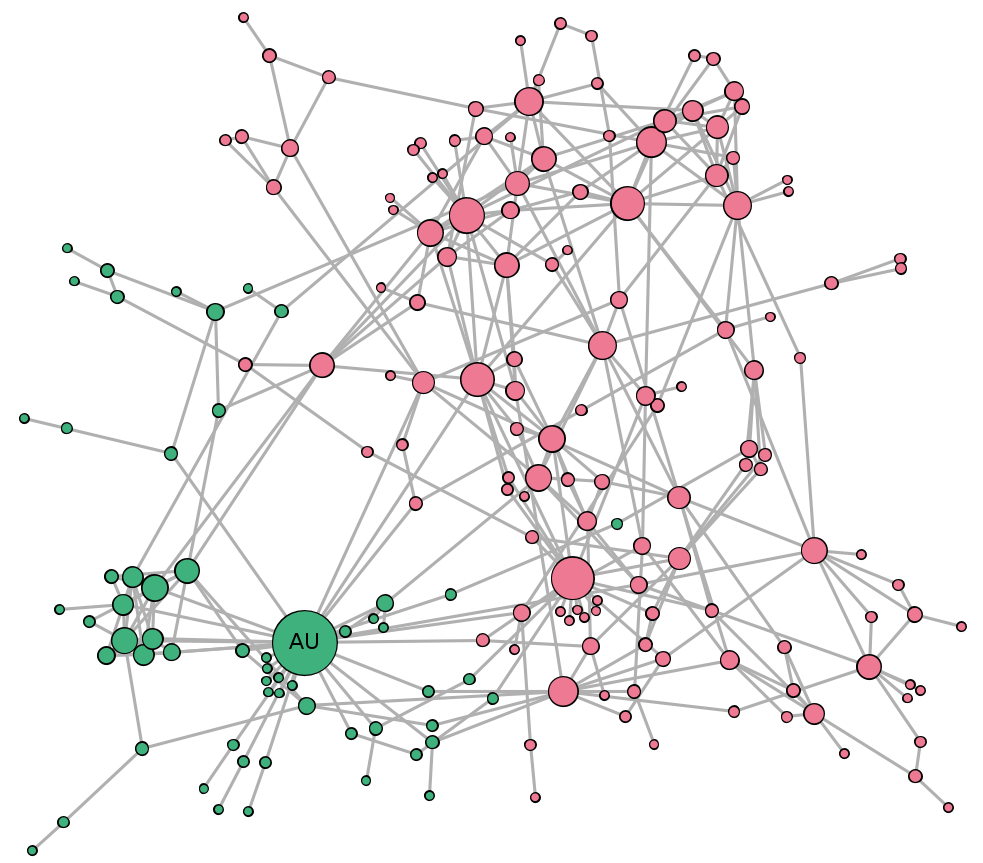}
    \caption{AN graphical representation. Size is proportional to the node degree and colors correspond to communities detected by the first cut of a Girvan-Newman algorithm. A clear fracture between the close supporters of Alvaro Uribe and the rest of the political scene may seen. }
    \label{fig:AN}
\end{figure}

The complete AN and WN are displayed in Figures \ref{fig:AN} and \ref{fig:WN}, respectively. The position of the nodes is calculated using a ``force based'' layout algorithm \cite{jacomy2011forceatlas2} which iteratively attracts linked nodes and  repels non-attached nodes. 

\subsubsection{AN Community Analysis}
The most notable characteristic of the AN  is how the node corresponding to Alvaro Uribe, despite having the high centrality values shown in Table \ref{tab:my_label},  distances itself from most other nodes with which it does not share  edges. 

The polarizing tendency is confirmed via community analysis using the  Girvan-Newman (GN) algorithm \cite{girvan2002community}. GN calculates the number of shortest paths between all nodes passing through each edge and removes the  edge with the highest count. When an edge removal makes the graph disconnected  the resulting connected components are considered communities and the process repeats for each component. In  AN we consider only the graph's first partition.

The first cut of AN results in two communities containing $69.27\%$ and $30.73\%$ of the nodes. The smaller community (shown in Figure \ref{fig:AN}) centers around Alvaro Uribe and contains many members of his party's inneokr circle including Colombia's current president Ivan Duque. 

The algebraic connectivity, also known as the the Fiedler eigenvalue  \cite{fiedler1989laplacian,chung1997spectral}, of a graph $G$ may be used  to partition the  graph partition. The Fiedler vector is the eigenvector corresponding to the second   smallest eigenvalue of the Laplacian matrix $L$  of $G$ where
\begin{equation}
 L=D-A,
\end{equation}
$A$ corresponds to the adjacency matrix of $G$ and $D$ to the degree matrix of $A$: 
\begin{equation} D_{i,j}:= \left\{
\begin{matrix}
deg(v_{i}) & \mbox{if}          & i=j\\
  0        & \mbox{otherwise.}  &
\end{matrix}
\right.
\end{equation}  

The sign (positive or negative) of each value of the Fiedler vector can be used to partition graph $G$. For the AN this results in a similar separation with only 5 nodes migrating from the largest to the smallest community. The proportional number of nodes remain similar   resulting in  $64.58\%$ and $35.42\%$ of nodes in each community.

\begin{figure}[t]
    \centering
\includegraphics[width=0.45\textwidth]{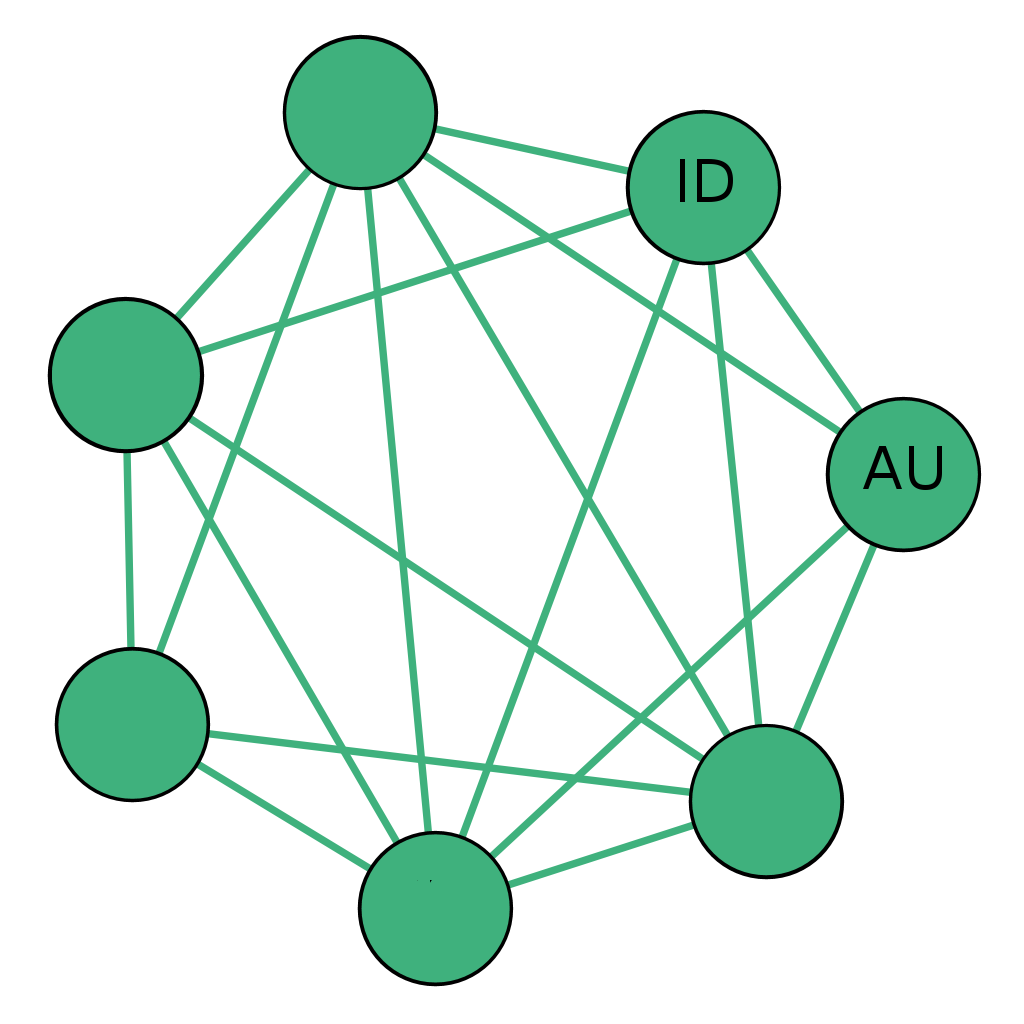}
    \caption{ There are three maximum cliques consisting of 5 nodes in the AN. All nodes correspond to members of the CD who are or have been members of the congress including former president Alvaro Uribe (AU) and current president Ivan Duque (ID).  }
    \label{fig:CliquesUribe}
\end{figure}

In both the GN and the Fiedler vector partitions, the community containing Alvaro Uribe also contains the three maximum cliques (subgraphs in which every node is connected to every other node).  The cliques are formed by 5 nodes each corresponding to current or former members of congress of the CD (including Alvaro Uribe and current President Ivan Duque).  As seen in Figure \ref{fig:CliquesUribe}, these cliques have an overlap of 4 nodes (equivalent to $80\%$  of the clique's nodes) between any pair of cliques indicating a strong and well connected inner network community  within the AN. 

\subsubsection{WN Community Analysis}

Unlike the AN, the WN does not present any easily detectable communities. The network revolves around Juan Manuel Santos and Alvaro Uribe, as predicted by the centrality values, but there is a dense network of shared vertices (people who have had work relationships with both)  which  bind them into one large community. 

\begin{figure}[t]
    \centering
        \includegraphics[width=0.75\textwidth]{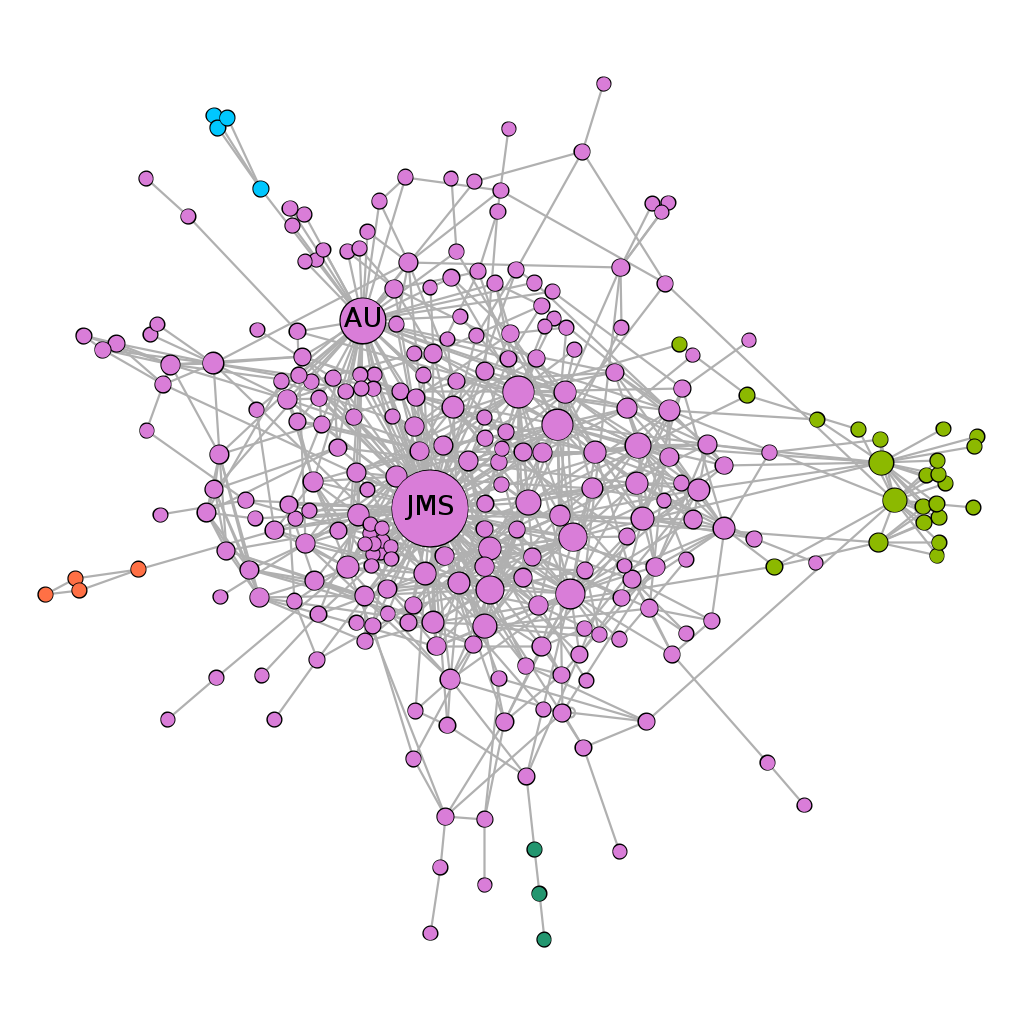}\
    \caption{WN graphical representation. Node colors correspond to the first five communities detected by the  Girvan-Newman algorithm. }
    \label{fig:WN}
\end{figure}

The first three cuts of GN correspond to very small marginal communities ($1.51\%$,$1.51\%$ and $1.13\%$ of all the nodes) connected by single edges to the main component.  It then separates a sizeable community  ( $8.68\%$ of the nodes) seen in the right side of Figure \ref{fig:WN}.  This community is formed mostly by politicians and technocrats who worked with Gustavo Petro, a prominent left-wing opposition leader, during his tenure as Bogota Mayor (2012-2015). 

\begin{figure}[t]
    \centering
\includegraphics[width=0.65\textwidth]{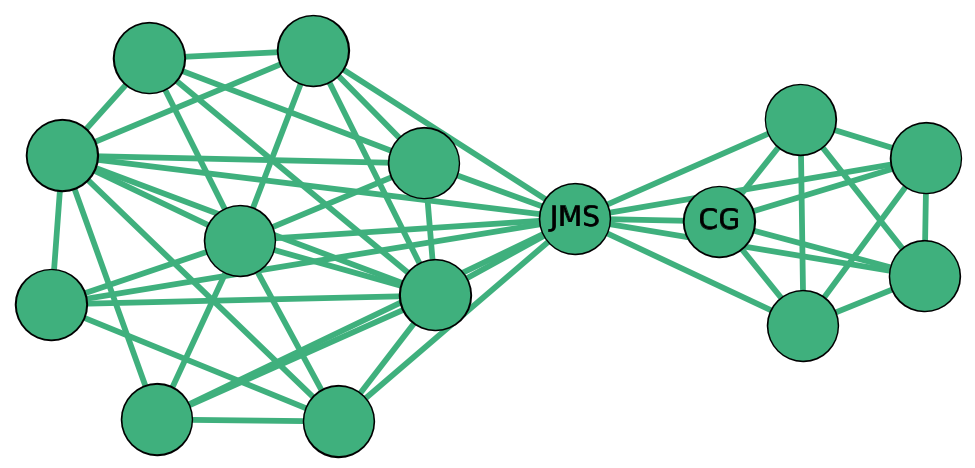}
    \caption{ There are 5 maximum cliques consisting of 6 nodes in the WN. Two communities bridged by former President Juan Manuel Santos (JMS) may be seen: A community formed by members of the Havana Peace Accords negotiating teams  (left) and a clique formed by members of the cabinet of former President Cesar Gaviria (CG, right). }
    \label{fig:CliquesSantos}
\end{figure}

The WN has 5 maximum cliques containing 6 nodes each.  The cliques form two distinct communities bridged by Juan Manuel Santos who is a member of 4 of them (see Figure \ref{fig:CliquesSantos}). The communities correspond to two distinct time periods: the negotiating team for the Havana Peace Accords (2012-2016)  and members of former President Cesar Gaviria's cabinet (1990-1994).  Unlike the maximal cliques of the AN, the WN is formed by representatives of various parties and sectors.

Whereas  AN is formed by two easily separable communities,  WN forms a tightly knit network without any strong separations. This would lead to think that, unlike the political arena,  in a a state as large and centralized as Colombia it is necessary through time  to draw people from all  political tendencies to keep the machinery moving.

\section{Conclusions}

A network-based approach to populism is necessary because it brings nuance to both the overly structural analysis and to the pairing of influential leader/faithful followers that so often is the backbone of the studies in populism. The massive popular support of a populist leader becomes the social force that allows eroding democratic institutions and mechanisms under his will, which is the most notorious danger of populism. That explains the interest in understanding both the leader’s followers, and the leader’s means to obtain support, by analyzing voting dispositions, discourse, ideology, and even personality traits in both voters and the leader. Additionally, studies of populist regimes also seek for explanations in structural terms: what are the social and economic conditions that allow a populist movement to thrive. These are all fundamental questions. However, they oscillate between a micro and a macro level. The network approach allows us to address the meso level by rendering visible the paths that connect the leader and the voters. These paths are made of politicians, advisors, and high-rank State officials that cement and increase or erode and diminish a populist leader's power. 

A characteristic of populist leaders is their presentation as outsiders. Uribe, launching his presidential campaign as an independent candidate, was not an exception. But just because someone presents him or herself as a political outsider does not mean that it is not playing the political game of forging alliances. Mapping the political landscape of Colombia allowed us to see the extent of Uribe’s partnerships and the uniqueness of his connections in comparison to other former Presidents. They are not nearly as interlinked as him and correspondingly their political presence is more marginal. Populist leaders need not only charisma, discourse, or ideology to get and maintain political power but some good old networking.

\section{Future research}

The data provided by the news website takes into account the most notable figures in Colombian politics based on their significance in a given news cycle, which makes it necessary to explain to the readers who the person is. Therefore, it is essential to build a database that includes not only the notable figures but also the ones that are less visible and equally powerful. This database would require collecting information about all the members both from the Congress and from the top Government positions. Additionally, the network approach to populism would benefit from tracing the changes in the alliances throughout the emergence and decline of a populist regime. Our long-term goal is to use the network morphology to create a populism indicator by  strengthening the database and incorporating a historical perspective in the analysis.

\bibliographystyle{plain}
\bibliography{refs}
%








\end{document}